\RequirePackage{fix-cm}
\documentclass[smallextended]{svjour3}       
\smartqed  
\usepackage{cite} 
\usepackage{graphicx}
\usepackage{amsmath}
\usepackage{amsfonts}
\newcommand{\td}{\mathrm{d}}
\newcommand{\w}{\omega}
\newcommand{\reef}[1]{(\ref{#1})}
\newcommand{\be}{\begin{equation}}
\newcommand{\ee}{\end{equation}}
\newcommand{\bal}{\begin{aligned}}
\newcommand{\eal}{\end{aligned}}

\def\d{\delta}
\def\e{\epsilon}           

\def\h{\eta}

\def\m{\mu}
\def\n{\nu}
  \def\w{\omega}

\def\tt{\tilde{\theta}}

\def\6{\partial}

\begin{document}

\title{Lovelock theory and the AdS/CFT correspondence
}


\author{Xi\'an O. Camanho \and Jos\'e D. Edelstein \and Jos\'e M. S\'anchez de Santos}


\institute{X. O. Camanho \at Department of Particle Physics and IGFAE, University of Santiago de Compostela, E-15782 Santiago de Compostela, Spain.\\
\email{xian.otero@rai.usc.es} \and
J. D. Edelstein \at Department of Particle Physics and IGFAE, University of Santiago de Compostela, E-15782 Santiago de Compostela, Spain.\\
Centro de Estudios Cient\'\i ficos (CECs), Valdivia, Chile.\\
\email{jose.edelstein@usc.es} \and
J. M. S\'anchez de Santos \at Department of Particle Physics and IGFAE, University of Santiago de Compostela, E-15782 Santiago de Compostela, Spain.\\
\email{josemanuel.sanchez.desantos@usc.es}
}

\date{Received: date / Accepted: date}

\maketitle

\begin{abstract}
Lovelock theory is the natural extension of general relativity to higher dimensions. It can be also thought of as a toy model for ghost-free higher curvature corrections in gravitational theories. It admits a family of AdS vacua, which provides an appealing arena to explore different holographic aspects in a broader setup within the context of the AdS/CFT correspondence. We will elaborate on these features and review previous work concerning the constraints that Lovelock theory entails on the CFT parameters when imposing conditions like unitarity, positivity of the energy or causality. 

\end{abstract}

\section{Introduction}
\label{intro}

Lovelock theories are the natural extension of the general relativity theory of gravity given by the Einstein-Hilbert action to higher dimensions and higher curvature interactions. The equations of motion do not involve terms with more than two derivatives of the metric, avoiding the appearance of ghosts
 \cite{Lovelock1971,Zwiebach,Zumino1986}. 

Much work has been done on the main properties of Lovelock gravity due to their interest as models where our knowledge of gravity can be tested and extended. For example, the vacua structure, the existence and properties of black holes such as their mass, entropy and thermodynamics, the gravitational phase transitions, the cosmological implications, etc. have been the object of an important amount of literature during the last years.

Nevertheless, the main motivation for this review article comes from the AdS/CFT correspondence, famously conjectured by Juan Maldacena some 15 years ago \cite{Maldacena}. This is nowadays well-stablished as a duality between quantum gravity theories in AdS space-times and conformal field theories living at the boundary. It is in that sense that the correspondence is dubbed customarily as the {\it holographic duality}. Originally formulated for $5$-dimensional AdS and $4$-dimensional CFT, lots of evidence accumulated over the years pointing towards its validity in higher and lower dimensions.

Lovelock theories have a rich structure of AdS vacua, which should be in correspondence with a similarly rich structure of higher dimensional CFTs. It is worth recalling at this point, however, that little is known\footnote{Parallel to this is the fact that Lovelock gravity might not be a consistent low energy truncation of any point in the moduli space of a putative UV complete (such as, for instance, M-) theory. In that respect, the relevance of these vacua is not {\it a priori} guaranteed.} about these higher dimensional CFTs. Not even their existence is clear. It has been argued that, in the supersymmetric case, there can be non trivial unitary CFTs in six dimensions, whose duals are seven dimensional gravity theories. Lovelock theories provide a useful framework to unravel some of the properties of higher dimensional CFTs, and also to test our understanding of the holographic duality when higher curvature terms come into play from the gravity side.

The subject is vast and it is far from our aim to cover it all. At those points where we consider that our presentation reduces to a bird's eye view, we will suggest further material where the interested reader can find more detailed explanations.

The article is organized as follows: we present the main features of Lovelock gravity using the first order formalism in section \ref{lovelock}. In section \ref{adscft} we review how  constraints on the CFT parameters are obtainted by holographically computing the two-point and three-point functions of the stress-energy tensor. The constraints come from unitarity and positivity of the energy. Section \ref{causality} is devoted to the analysis of possible causality violations by considering the scattering of gravitons against shock waves propagating in a Lovelock AdS background. The results are in agreement  with those of section \ref{adscft} and also with the ones obtained by a similar calculation performed in the perturbed black hole background dual to a thermal field theory.

In section \ref{final} we present the conclusions, add some final comments, review recent developments in the subject and give some possible directions for future work.

\section{Lovelock theory}
\label{lovelock}

Some four decades ago, David Lovelock derived a formal expression for the most general, symmetric and conserved tensor which is quasi-linear in the second derivatives of the metric without any higher derivatives in arbitrary space-time dimensionality \cite{Lovelock1971}. They provide an interesting playground to explore the effect of higher curvature terms in the framework of the AdS/CFT correspondence. Very recent reviews on general aspects of this theory include \cite{Padmanabhan2013,Edelstein2013}.

\subsection{Preliminaries}

For the sake of making progress in the study of Lovelock theory, it is convenient to use differential forms and the exterior algebra (see, for instance, \cite{Eguchi1980,Willison2004,Zanelli2005}). Instead of the metric and affine connection, we will be referring to orthonormal frames (or {\it vielbein}) and spin connection (or connection 1-form) \cite{Regge}. This formalism will make our expressions much more compact and also the manipulations much easier. The vielbein is a non-coordinate basis which provides an orthonormal basis for the tangent space at each point on the manifold,
\begin{equation}
g_{\m\n}\, dx^\m \otimes dx^\n = \h_{ab}\, e^a \otimes e^b ~,
\end{equation}
where $\h_{ab}$ is the $d$-dimensional Minkowski metric with $(-1,1,\ldots,1)$ signature. The Latin indices $\{a,b,\ldots\}$ are \textit{flat} or \textit{tangent space indices}, while the Greek ones $\{\m,\n,\ldots\}$ are  \textit{curved} or \textit{spacetime indices}. In some cases we will also distinguish spacelike $\{i,j,\ldots\}$ from timelike ones. The vielbein are $d$ 1-forms, 
\begin{equation}
e^{a}=e^{a}_{\m}\,\td x^{\m} ~,
\end{equation}
that we may use in order to rewrite the metric as
\begin{equation}
g_{\m\n}=\eta_{ab}\,e^{a}_{\m}\,e^{b}_{\n} ~.
\end{equation}
We also need to introduce the metric compatible (antisymmetric) connection 1-form $\omega^a_{~b}$ that is necessary in order to deal with tensor valued differential forms. In addition to the usual exterior derivative, $\td$, we define the {\it covariant} exterior derivative, $D$, that reduces to the former when applied to a scalar valued form. For a general $(p,q)$-tensor valued form
\begin{equation}
DV^{a_1\cdots a_p}_{b_1 \cdots b_q} := \td V^{a_1\cdots a_p}_{b_1 \cdots b_q}+\sum_{i=1}^p \w^{a_i}_{\ c}\wedge V^{a_1\cdots c \cdots a_p}_{b_1 \cdots b_q}-\sum_{j=1}^q \w^{d}_{\ b_j}\wedge V^{a_1 \cdots a_p}_{b_1 \cdots d \cdots b_q} ~.
\end{equation}
We can in this way define the torsion and curvature 2-forms as derivatives of, respectively, the vielbein and the spin connection
\begin{eqnarray}
T^a &:=& De^a ~, \\ [0.5em]
R^{ab} &:=& \td\w^{ab}+\w^{a}_{\ c}\wedge \w^{cb}=\frac{1}{2} R_{~b\mu\nu}^{a}\; dx^{\mu} \wedge dx^{\nu} ~,
\end{eqnarray}
known as the Cartan structure equations. The covariant derivative of Cartan's equations give the Bianchi identities
\begin{equation}
DT^a = R^a_{\ b}\wedge e^b ~, \qquad\qquad\qquad DR^{ab} = 0 ~.
\end{equation}
We will consider a sector of Lovelock theory where the torsion vanishes. This is not the most general situation, but it will suffice the purpose of this article. In the absence of torsion, the spin connection is not independent from the metric and coincides with the Levi-Civita connection,
\begin{equation}
\omega^a_{~b} = e^a_\mu e^\nu_b \Gamma^{\mu}_{~\nu\rho}\,\td x^{\rho} ~.
\end{equation}
In GR the torsion tensor is constrained to vanish. When this constraint is not imposed, we have the Einstein-Cartan theories. These are very important when considering spinor fields as these generally source the spin connection.

For later use, it is convenient to introduce some further notation:
\begin{eqnarray}
R^{a_1 \ldots a_{2n}} &:=& R^{a_1 a_2}\wedge\ldots \wedge R^{a_{2n-1} a_{2n}} ~, \\ [0.5em]
e^{a_1 \ldots a_n} &:=& e^{a_1}\wedge \ldots \wedge e^{a_n} ~.
\end{eqnarray}
We will also use the antisymmetric tensor $\e_{a_1 \ldots a_d}$ when writing down and manipulating the Lovelock lagrangian and the derived equations of motion. It is antisymmetric on any pair of indices with $\e_{123\ldots d}=+1$. Some times, in order to deal with more compact expressions, we will write scalars constructed with the antisymmetric tensor, such as
\begin{equation}
\e\!\left[\psi\right] = \e_{a_1 \ldots a_d} \psi^{a_1 \ldots a_d} ~.
\end{equation}
%

\subsection{The Lovelock action and its Euler-Lagrange equations}

The action of Lovelock theory is given by
\begin{equation}
\mathcal{I} =\frac{1}{16\pi G_N (d-3)!}\, \sum_{k=0}^{K} {\frac{c_k}{d-2k}} \mathop\int \mathcal{L}_{k} ~,
\label{LLaction}
\end{equation}
$G_N$ being the Newton constant in $d$ spacetime dimensions. $\{c_k\}$ is a set of couplings with length dimensions $L^{2(k-1)}$, $L$ being a length scale related to the cosmological constant, while $K$ is a positive integer,
\begin{equation}
K\leq \left[\frac{d-1}{2}\right] ~,
\label{maximalK}
\end{equation}
labeling the highest non-vanishing coefficient, {\it i.e.}, $c_{k>K} = 0$. $\mathcal{L}_{k}$ is the exterior product of $k$ curvature 2-forms with the required number of vielbein, $e^a$, to construct a $d$-form,
\begin{equation}
\mathcal{L}_{k} = \e\!\left[R^k e^{d-2k}\right] = \epsilon_{a_1 \ldots a_{d}}\; R^{a_1 \ldots a_{2k}} \wedge e^{a_{2k+1} \ldots a_d} ~.
\label{LLaction-k}
\end{equation}
The zeroth and first term in (\ref{LLaction}) correspond, respectively, to the cosmological term and the Einstein-Hilbert action. It is fairly easy to see that $c_0 = L^{-2}$ and $c_1 = 1$ correspond to the usual normalization of these terms, the cosmological constant having the customary negative value $2\hat\Lambda = - (d-1)(d-2)/L^2$. Either a positive ($c_0 = -L^{-2}$) or a vanishing ($c_0 = 0$) cosmological constants can be easily incorporated as well. The first non-trivial Lovelock term,
\begin{equation}
\mathcal{L}_{2} = R^2 - 4 R_{\mu\nu} R^{\mu\nu} + R_{\mu\nu\rho\sigma} R^{\mu\nu\rho\sigma} ~,
\label{LGB}
\end{equation}
contributes just for dimensions larger than four \cite{Lanczos}, and corresponds to the Lanczos-Gauss-Bonnet (LGB) coupling $c_2=\lambda L^2$. The Kaluza-Klein reduction of LGB theory and its corresponding cosmological scenarios have been first considered in \cite{Madore}. We will also discuss below the case of cubic Lovelock theory, whose contribution reads
\begin{eqnarray}
\mathcal{L}_{3} &=& R^3\! +\! 3 R R^{\mu\nu\alpha\beta} R_{\alpha\beta\mu\nu}\! -\! 12 R R^{\mu\nu} R_{\mu\nu}\! +\! 24 R^{\mu\nu\alpha\beta} R_{\alpha\mu} R_{\beta\nu}\! +\! 16 R^{\mu\nu} R_{\nu\alpha} R_{\mu}^{~\alpha} \nonumber \\ [0.5em]
& & + 24 R^{\mu\nu\alpha\beta} R_{\alpha\beta\nu\rho} R_{\mu}^{~\rho} + 8 R_{~~\alpha\rho}^{\mu\nu} R_{~~\nu\sigma}^{\alpha\beta} R_{~~\mu\beta}^{\rho\sigma} + 2 R_{\alpha\beta\rho\sigma} R^{\mu\nu\alpha\beta} R_{~~\mu\nu}^{\rho\sigma} ~.
\label{LL3}
\end{eqnarray}
and the corresponding coupling is $\mu = 3 c_3/L^{4}$. This latter expression is cumbersome enough to shed light on the reasons why it is much more convenient to work with expressions like \reef{LLaction-k} rather than the usual tensorial formalism.

In first order formalism, we shall consider the vielbein and the spin connection as independent variables. We then have two equations of motion, one for each {\it field}. Varying the action with respect to the spin connection 1-form results in
\begin{eqnarray}
\delta_{\w}\mathcal{L}_{k}&=& k\,\e\!\left[D(\delta\w)R^{k-1}e^{d-2k}\right] \nonumber\\ [0.5em]
&=&k\,\td \e\!\left[\d\w R^{k-1} e^{d-2k}\right] - k(d-2k)\,\e\!\left[\d\w T R^{k-1}e^{d-2k-1}\right] ~,
\label{spineq}
\end{eqnarray}
where we have used that $\delta_{\w}R^{ab}=D(\delta\w^{ab})$, integration by parts (we use the technology of exterior algebra and treat exterior covariant derivatives as normal derivatives inside the brackets), and the Bianchi identity $DR^{ab}=0$. The first term in the above variation is a total derivative and does not contribute to the equations of motion whereas the second is proportional to the torsion. We may safely restrict to the torsionless sector, allowing us to compare our results with those coming from the tensorial formalism based on the metric.

Even though the first term in \reef{spineq} is irrelevant for the matter of discussing solutions to the Lovelock equations, it contributes to the variation of the action in such a way that we need to include boundary terms analogous to that of Gibbons and Hawking \cite{Gibbons1977} for General Relativity. These terms precisely cancel the previous boundary contributions so that the Lovelock action defines a well posed variational problem. In the same way as the Lovelock terms are the (dimensionally continued) Euler densities for manifolds in $2k$ dimensions, the corresponding boundary terms appear in the generalization of the Gauss-Bonnet theorem to manifolds with boundaries \cite{Myers1987}
\begin{equation}
\mathcal{Q}_k = k \int_0^1\!\! d\xi \ \e\!\left[\theta\,\mathfrak{F}_\xi^{k-1} e^{d-2k}\right] ~,
\label{LLboundary}
\end{equation}
$\theta^{ab}$ is the second fundamental form associated to the extrinsic curvature, and
\begin{equation}
\mathfrak{F}_\xi^{ab} := R^{ab} + (\xi^2 - 1)\, \theta_{~e}^{a} \wedge \theta^{eb} ~.
\label{curlyF}
\end{equation}
These terms play also a central r\^ole in deriving junction conditions, such as the Israel conditions in General Relativity \cite{Israel1967}, for these gravity theories \cite{Gravanis2003a}. These matching equations have been exploited extensively in \cite{CEGG1,CEGG2} for the sake of finding distributional metrics, in the absence of matter, that allow to describe new types of phase transitions between different branches of Lovelock theory.

The second equation of motion is obtained by varying the action with respect to the vielbein. It can be casted into the form
\begin{equation}
\mathcal{E}_a := \epsilon_{a a_1 \ldots a_{d-1}}\;c_K\, \mathcal{F}_{(1)}^{a_1 a_2} \wedge \cdots \wedge \mathcal{F}_{(K)}^{a_{2K-1} a_{2K}} \wedge e^{a_{2K+1}\ldots a_{d-1}} = 0 ~,
\label{eqlambda}
\end{equation}
where $\mathcal{F}_{(i)}^{a b} := R^{a b} - \Lambda_i\, e^a \wedge e^b$, $\Lambda_i$ being a function of the Lovelock couplings. This expression involves just the curvature 2-form and no extra covariant derivatives, making explicit the two derivative character of the Lovelock equations of motion. Also, for the critical dimension $d=2k$, the $k^{th}$ term contribution to the equations of motion vanishes. In our approach this is simply due to the absence of vielbein fields in the corresponding action term, thus yielding zero upon variation. More generally, the integral of that term becomes a topological invariant, the Euler number for that particular dimension. In dimensions lower than the critical one the corresponding Lovelock term exactly vanishes and we are led to the restriction (\ref{maximalK}).

\subsection{Constant curvature vacuum solutions}

It is transparent from (\ref{eqlambda}) that, in principle, this theory admits $K$ constant curvature {\it vacuum} solutions,
\begin{equation}
\mathcal{F}_{(i)}^{a b} = R^{a b} - \Lambda_i\, e^a \wedge e^b = 0 ~.
\end{equation}
Indeed, inserting $R^{ab}=\Lambda\,e^a \wedge e^b$ in (\ref{eqlambda}), one finds that the $K$ different effective {\it cosmological constants} are the (real) solutions of the $K^{th}$ order {\it characteristic} polynomial
\begin{equation}
\Upsilon[\Lambda] := \sum_{k=0}^{K} c_k\, \Lambda^k = c_K \prod_{i=1}^K \left( \Lambda - \Lambda_i\right) =0 ~,
\label{cc-algebraic}
\end{equation} 
each one corresponding to a different {\it vacuum}, positive, negative or zero for dS, AdS and flat spacetimes. The effective cosmological constants correspond to the (inverse squared of the) possible radii of these (A)dS spaces and should not be confused with the bare cosmological constant, $\hat\Lambda$ appearing in the action. The theory will have degenerate behavior whenever two or more effective cosmological constants coincide. This is captured by the discriminant,
\begin{equation}
\Delta = \prod_{i < j}^K (\Lambda_i - \Lambda_j)^2 ~,
\label{discriminant}
\end{equation}
that vanishes in a certain locus of the parameter space given by the coupling constants of Lovelock theory where some special features arise. The discriminant can be written as well in terms of the first derivative of the Lovelock polynomial, $\Upsilon$, as
\begin{equation}
\Delta = \frac{1}{c_K^K}\prod_{i=1}^K |\Upsilon'[\Lambda_i]| ~.
\end{equation}
As we move forward it will become clear the preeminent r\^ole played by this polynomial in the most diverse situations. Another property of any degenerate vacuum is the absence of linearized gravitational degrees of freedom about it. The equations of motion for a metric perturbation around a given vacuum, $\Lambda_1$, are easily obtained from the perturbation of the curvature
\begin{equation}
R^{ab} = \Lambda_1 e^{ab} + \delta_{\!g}R^{ab} ~,
\end{equation}
yielding, at linear level,
\begin{equation}
\mathcal{E}_a = \Upsilon'[\Lambda_1]\,\e_{a a_1 \ldots a_{d-1}} \delta_{\!g} R^{a_1 a_2} \wedge e^{a_3 \ldots a_{d-1}} ~;
\end{equation}
thus, it is exactly zero as long as the first derivative of $\Upsilon$ vanishes for a degenerate vacua, 
\begin{equation}
\Upsilon'[\Lambda_1]=c_K\prod_{i\neq 1}(\Lambda_1-\Lambda_i)=0 ~.
\end{equation}
Moreover, it is easy to verify that the equations of motion around a non-degenerate vacuum are exactly those of the Einstein-Hilbert gravity, multiplied by a global factor proportional to $\Upsilon'[\Lambda_1]$. The propagator of the graviton corresponding to the vacuum $\Lambda_1$ is then proportional to $\Upsilon'[\Lambda_1]$, in such a way that when $\Upsilon'[\Lambda_1]<0$, it has the opposite sign with respect to the Einstein-Hilbert case and, thus, the graviton becomes a ghost. This generalizes the observation first done by Boulware and Deser \cite{Boulware1985a} in the context of LGB gravity. Thus, a given vacuum of Lovelock gravity, $\Lambda_\star$, must satisfy
\begin{equation}
\Upsilon'[\Lambda_\star] > 0 ~,
\end{equation}
in order to host gravitons propagating with the right sign of the kinetic term.

\subsection{Shock wave solutions}

We are going to use shock wave backgrounds of the Lovelock theory for holographic applications. These are very interesting solutions with a particular structure that make them exact solutions on any gravity theory. In fact, they are not corrected when higher curvature corrections are included and were shown to be exact solutions of string theory \cite{Horowitz1999}. A shock wave propagating on AdS along the radial direction has the form
\begin{equation}
ds^2_{\rm AdS,sw} = ds^2_{\rm AdS} + f(u)\, \varpi(\mathtt{x},z)\, du^2 ~,
\label{sw}
\end{equation}
where we have defined light-cone coordinates $u = t + x^{d-1}$ and $v = t - x^{d-1}$, $\mathtt{x}$ is the $d-3$ vector whose components are $x^a$, $a=2,\ldots,d-2$, and $f(u)$ is an arbitrary distribution with support in $u=0$, which we will identify later as a Dirac delta function, $f(u)=\delta(u)$, for simplicity. $ds^2_{\rm AdS}$ is the AdS metric in Poincar\'e coordinates, which is a solution of the Einstein equations with cosmological constant $\Lambda_\star$,
\begin{equation}
ds^2_{\rm AdS} = - \frac{1}{\Lambda_\star z^2}\, \left( {-du dv + d\mathtt{x}^2 + dz^2} \right) ~.
\end{equation}
The equation of motion for the shock wave profile is
\begin{equation}
2(d-3)\varpi+(d-6)z\,\partial_z \varpi-z^2(\nabla_\mathtt{x}^2 \varpi + \partial_z^2 \varpi)=0 ~.
\label{SWeq}
\end{equation}
This equation admits a few solutions, depending upon the assumptions made regarding the coordinate dependence of $\varpi$, whose r\^ole will be discussed in what follows:
\begin{eqnarray}
& & \varpi_1(z) = \varpi_0\, {z^{d-3}} ~, \qquad\qquad \varpi_2(z) =\frac{\varpi_0}{z^2} ~, \label{wsol2}\\ [0.5em]
& & \varpi_3(\mathtt{x},z) = \frac{\varpi_0\,z^{d-3}}{\left(z^2 + (\mathtt{x} - \mathtt{x_0})^2\right)^{d-2}} ~.
\label{wsol3}
\end{eqnarray}
More concretely, solutions $\varpi_1(z)$ and $\varpi_3(\mathtt{x},z)$  will be relevant for our discussions below, while $\varpi_2(z)$ is just a coordinate redefinition \cite{Hofman2008}.

\section{The AdS/CFT correspondence}
\label{adscft}

We will not attempt to review in detail such a vast subject in the present article. In turn, let us make a highly pragmatical construction limiting our presentation just to those features that we will need. Since we are going to work in the realm of Lovelock theory, whose UV completion is unknown, we will adopt a rough version of the AdS/CFT correspondence assuming that: 
\begin{quotation}
\noindent
(Quantum) gravity in AdS is dual to a CFT {\it living} at the boundary.
\end{quotation}
The interested reader may find convenient to dig into the classic review \cite{Aharony1999}. For applications of the AdS/CFT correspondence to strongly coupled phenomena in QCD-like theories, see \cite{Edelstein2006,Casalderrey2011}.

\subsection{Correlation functions}

The CFT dynamics is entirely given by the correlation functions of its gauge invariant local operators. The dynamical information of the AdS/CFT correspondence can then be embodied in the relation between the generating functional for CFT correlators and the string theory/quantum gravity in AdS partition function with appropriate boundary conditions \cite{GKP,Witten98}. In the case of a scalar field, $\phi(z, \mathbf x)$, for instance, we have
\begin{equation}
\left\langle \exp\left(i\int_\Sigma dx^\mu \phi_0(\mathbf x)\,\mathcal{O}(\mathbf x)\right)\right\rangle_{\rm CFT} = \mathcal{Z}_{\rm QG\,in\,AdS}\left[\phi(0,\mathbf x)=\phi_0(\mathbf x)\right] ~,
\label{AdSCFTgolden}
\end{equation}
where the asymptotic value (or boundary condition) of the scalar field in AdS acts as a source for the dual scalar operator $\mathcal{O}(\mathbf x)$. Taking into account the holographic statement establishing that the degrees of freedom of the CFT are confined to the boundary, we can only define local gauge invariant observables like $\mathcal{O}(\mathbf x)$ precisely at $z=0$.

The hamiltonian is realized in the CFT as the dilatation operator, in such a way that the energy in AdS corresponds to the conformal dimension in the dual CFT, these being related as 
\begin{equation}
m^2=\Delta(\Delta-d) ~.
\end{equation}
The same can be also generalized to operators with spin. In the CFT side the spectrum will always contain many operators with different spins and conformal dimensions. One of them is {\it universal} in the sense that it is present for any CFT and has some very specific properties. It is the stress-energy tensor and it is sourced in the dual picture by the boundary value of the graviton field. We shall restrict our discussion to purely gravitational theories which then amounts to the analysis of correlators involving the stress-energy tensor. In that case, we can compute the generating function as
\begin{equation}
\left\langle \exp \bigg( \int\!d\mathbf x ~\eta^{ab}(\mathbf x)\; T_{ab}(\mathbf x) \bigg) \right\rangle_{\!\rm CFT} = \mathcal{Z}_{\rm QG\,in\,AdS} \left[g_{ab}(0,\mathbf x) = \eta_{ab}(\mathbf x)\right] ~,
\end{equation}
where $\mathcal{Z}_{\rm QG\,in\,AdS}$ is the partition function of quantum gravity in AdS spacetime, the same that we use to discuss thermodynamic properties of black holes, integrated over all metrics $g_{\mu\nu} = g_{\mu\nu}(z,\mathbf x)$ satisfying the boundary condition $g_{ab}(0,\mathbf x) = \eta_{ab}(\mathbf x)$. In the saddle point approximation, this partition function can be computed as just the classical contribution,
\begin{equation}
\left\langle \exp \bigg( \int\!d\mathbf x ~\eta^{ab}(\mathbf x)\; T_{ab}(\mathbf x) \bigg) \right\rangle_{\!\rm CFT}\approx \exp \left( - \mathcal{I}_{\rm on-shell}[\eta_{ab}(\mathbf x)] \right) ~,
\label{goldrule}
\end{equation}
{\it i.e.}, the on-shell action for the (least action) classical solution with the given boundary conditions. From this expression, correlators of the stress-energy tensor can be obtained by performing functional derivatives of \reef{goldrule} with respect to the boundary metric. This, in turn, is simply given by considering gravitational fluctuations around an asymptotically AdS configuration of the theory. The bulk metric acts as a source for the stress-energy tensor in the boundary (and viceversa).

\subsection{CFT unitarity and 2-point functions}

The leading singularity of the $2$-point function of a CFT in $(d-1)$ dimensions, when canonical normalization has been adopted for the fundamental fields, is fully characterized by a single number, $C_T$, known as the central charge \cite{OsbornPetkou}
\begin{equation}
\langle T_{ab}(\mathbf x)\, T_{cd}(\mathbf 0)\rangle = \frac{C_T}{\mathbf x^{2(d-1)}}\;\mathcal{I}_{ab,cd}(\mathbf x) ~,
\label{TTcorrelator}
\end{equation}
where the index structure is given by
$$
\mathcal{I}_{ab,cd}(\mathbf x) = \frac12 \left( I_{ac}(\mathbf x)\, I_{bd}(\mathbf x) + I_{ad}(\mathbf x)\, I_{bc}(\mathbf x) - \frac1{d-1}\, \eta_{ab}\, \eta_{cd} \right) ~,
$$
with
$$
I_{ab}(\mathbf x) = \eta_{ab} - 2\,\frac{x_a\, x_b}{\mathbf x^2} ~.
$$
For instance, $C_T$ is proportional in a four dimensional CFT to the standard central charge $c$ that multiplies the (Weyl)$^2$ term in the trace anomaly, $C_T = 40\, c/\pi^4$. Following the prescription presented above, the central charge has been computed holographically  in \cite{Buchel2010a} for LGB in various dimensions and in \cite{Camanho2010d} for Lovelock theory. We shall expand this computation in the following.

Since we are interested in a two-point function, the action has to be computed to second order in the metric perturbation. It can be shown that, when the background metric is an AdS solution of the Lovelock action with cosmological constant, say $\Lambda_1$, then its second variation is proportional to that of the Einstein-Hilbert action, the proportionality being given by the derivative of the characteristic polynomial,
\begin{equation}
\delta^2 {\cal I} = \Upsilon'[\Lambda_1]\,\delta^2 {\cal I}_{EH} ~.
\label{desres}
\end{equation}
The proof goes as follows. In the first order formalism, the variation of the Lovelock action can be written as $\delta {\cal I} = \int\!{\cal E}_a\,\delta e^a$, where $\delta e^a$ is the variation of the vielbein one-form. Variations with respect to the spin connection vanish since they are proportional to the torsion. The Lovelock equations of motion (\ref{eqlambda}) are ${\cal E}_a=0$. For the second variation we have
\begin{equation}
\delta^2 {\cal I} = \int {\cal E}_a\,\delta^2 e^a + \frac{\partial {\cal E}_a}{\partial e^b}\,\delta e^b \wedge \delta e^a ~,
\end{equation}
where the first term vanishes when evaluated on-shell, $R^{ab}= \Lambda_1 e^{a} \wedge e^{b}$. The only non-vanishing contributions of the second term, in turn, come from the derivative acting on the ${\cal F}_{(1)}^{ab}$ factor,
\begin{equation}
\delta^2 {\cal I} = \prod_{k \neq 1}^K (\Lambda_1 - \Lambda_k)
\int \epsilon_{a a_1 \ldots a_{d-1}} \frac{\partial {\cal F}_{(1)}^{a_1a_2}}{\partial e^b} \wedge e^{a_3 \ldots a_{d-1}} \delta e^b \wedge \delta e^a ~.
\end{equation}
The pre-factor is easily seen to be $\Upsilon'[\Lambda_1]$, while the remaining integral is just the second variation of the Einstein-Hilbert action, leading to \reef{desres}. The holographic calculation of the central charge of the CFT dual to a Lovelock gravitational theory is then completely analogous to that of the Einstein-Hilbert case, provided the boundary terms and counterterms are properly taken into account \cite{Myers1987,Arutyunov1999}. Let us briefly review this calculation, following \cite{Liu1998}. For that purpose, consider the perturbation
\begin{equation}
g_{\mu \nu} = g^0_{\mu \nu} + h_{\mu \nu} ~,
\end{equation}
where the background metric $g^0_{\mu \nu}$ is the AdS metric in $d$ dimensions,
\begin{equation}
ds^2=g^0_{\mu \nu}\,dx^\mu dx^\nu = \frac{dz^2 + d\textbf{x}^2}{z^2} = \frac{dz^2 + \eta_{a b}\,dx^a dx^b}{z^2} ~,
\end{equation}
where we have considered, for simplicity, $\Lambda_1=-1$; restoring $\Lambda_1$ factors at any step is straightforward on dimensional grounds. The boundary at $z=0$ is flat, with metric $g_{ab}(0,\textbf{x}) = \eta_{ab}$. The background metric satisfies
\begin{equation}
R^0_{\mu \nu} = -(d-1)\, g^0_{\mu \nu} ~,
\end{equation}
where $R^0_{\mu \nu}$ is the background Ricci tensor in the expansion 
$R^{\mu \nu} = R^0_{\mu \nu} + R^1_{\mu \nu} + R^2_{\mu \nu} + \ldots$ The complete gravity action can be expanded in powers of $h_{\mu \nu}$. Up to total derivative contributions, which are cancelled by the boundary terms, the quadratic part is
\begin{eqnarray}
{\cal I}_2 &=& \frac{1}{4}\Upsilon'[\Lambda_1] \int\! dz\,d\textbf{x}\,\sqrt{g^0} \left[ D_\mu h D^\mu h - 2 D^\mu h D^\nu h_{\mu \nu} \right. \nonumber\\ [0.5em]
&& \qquad \left. + 2 D^\mu h^{\alpha \beta} D_\alpha h_{\mu \beta}- D^\mu h^{\alpha \beta} D_\mu h_{\alpha \beta} \right] ~,
\end{eqnarray}
where $h$ is the trace of the perturbation. Using the equations of motion, the action becomes a total derivative which can be reduced to an integral over the boundary. Terms without derivatives in $h_{\mu \nu}$ are cancelled by the action counterterm proportional to the boundary volume. With the appropriate choice of gauge in which $h_{00} = h_{0a} = 0$, the action can be finally written as
\begin{equation}
{\cal I}_2 = \frac{1}{4}\Upsilon'[\Lambda_1] \int\! d\textbf{x} ~z^{2-d} ~{\bar h}^{ab} \partial_z {\bar h}_{ab} ~,
\end{equation}
where ${\bar h}_{ab}$ is the traceless part of the perturbation. The generating functional for stress-energy tensor correlation functions is then the gravity action evaluated on-shell. This implies solving the Dirichlet problem for $h_{\mu \nu}(z,\textbf{x})$ consisting on the Einstein equation at linear order
\begin{equation}
R^1_{\mu \nu} = -(d-1)\,h_{\mu \nu} ~,
\end{equation}
with the boundary conditions:
\begin{equation}
z^2 h_{ab}(0,\textbf{x}) = \mathtt{h}_{ab}(\textbf{x}) \qquad h_{za}(0,\textbf{x}) = h_{zz}(0,\textbf{x}) = 0 ~.
\end{equation}
The solution can be written as:
\begin{equation}
h_{\mu \nu}(z,\textbf{x}) = \int\! d\textbf{y}\; {\cal K}_{\mu \nu, ab}(z,\textbf{x},\textbf{y})\;\mathtt{h}_{ab} (\textbf{y}) ~,
\end{equation}
in terms of the bulk to boundary Green's function
\begin{equation}
{\cal K}_{\mu \nu, ab}(z,\textbf{x},\textbf{y}) = \kappa_d \frac {z^{d-3}}{\big(z^2 + {\left|\textbf{x}-\textbf{y}\right|}^2\big)^{d-1}}~ {\cal I}_{\mu \nu, ab}(\mathbf{x} - \mathbf{y})\; \mathtt{h}_{ab} (\textbf{y}) ~.
\end{equation}
Here $\kappa_d$ is a normalization constant ensuring that the propagator becomes a $\delta$ function at $z=0$,
\begin{equation}
\kappa_d = \frac{d}{d-2} \frac{\Gamma(d-1)}{\pi^{\frac{d-1}{2}} \Gamma(\frac{d-1}{2})} ~.
\end{equation}
Inserting this result into the quadratic action, we obtain
\begin{equation}
{\cal I}_2=\frac{d(d-1)}{4(d-2)} \frac{\Gamma(d-1)}{\pi^{\frac{d-1}{2}} \Gamma(\frac{d-1}{2})}\Upsilon'[\Lambda_1] \int\! d\textbf{x}\,d\textbf{y} ~\frac{ \mathtt{h}_{ab}(\textbf{x})~{\cal I}_{ab,cd}(\textbf{x} - \textbf{y})~\mathtt{h}_{cd}(\textbf{y})}{{\left|\textbf{x} -\textbf{y}\right|^{2(d-1)}}} ~,
\end{equation}
from where we read the central charge. Restoring factors of $\Lambda_1$ on dimensional grounds this yields
\begin{equation}
C_T = \frac{d}{2(d-2)}\frac{\Gamma (d)}{\pi^{\frac{d-1}{2} \Gamma(\frac{d-1}{2})}} \frac{\Upsilon'[\Lambda_1]}{(-\Lambda_1)^{\frac{d-2}{2}}} ~.
\end{equation}
Then, holography teaches us that the requirement that Boulware-Deser instabilities (gravitons propagating with kinetic terms of the wrong sign) are absent in Lovelock gravity, $\Upsilon'[\Lambda_1] > 0$, is equivalent to the positivity of the central charge; {\it i.e.}, to the condition of unitarity of the corresponding dual CFT.

\subsection{Three-point function and conformal collider physics}
\label{3pcorr}

The form of the $3$-point function of the stress-tensor in a $(d-1)$-dimensional conformal field theory is highly constrained. It was shown in \cite{OsbornPetkou,Erdmenger1997} that it can always be written in the form
\begin{equation}
\langle T_{ab}(\mathbf x)\,T_{cd}(\mathbf y)\,T_{ef}(\mathbf z)\rangle = 
\frac{\mathcal A\, \mathcal I^{(1)}_{ab,cd,ef}+\mathcal B\, \mathcal I^{(2)}_{ab,cd,ef}+\mathcal C\, \mathcal I^{(3)}_{ab,cd,ef}}{|\mathbf x - \mathbf y|^{d-1}\,|\mathbf y - \mathbf z|^{d-1}\,|\mathbf z - \mathbf x|^{d-1}} ~,
\end{equation}
where the form of the tensor structures $\mathcal I^{(i)}_{ab,cd,ef}$ will be irrelevant for us here. Energy conservation also implies a relation between the central charge $C_T$ appearing in the $2$-point function, and the parameters $\mathcal A,\mathcal B,\mathcal C$, namely
\begin{equation}
C_T = \frac{\pi^{\frac{d-1}{2}}}{\Gamma\left[\frac{d-1}2\right]}\,\frac{(d-2)(d+1)\mathcal{A}-2\mathcal{B}-4 d \,\mathcal{C}}{(d-1)(d+1)} ~.
\end{equation}
Since we have already computed $C_T$ in the previous section, we are left with two independent parameters to be calculated.

A convenient parameterization of the $3$-point function of the stress-tensor was introduced in \cite{Hofman2008}, where a {\it gedanken} collision experiment is considered in an arbitrary CFT$_{d-1}$. One wants to measure the total energy flux per unit angle deposited in calorimeters distributed around the collision region,
\begin{equation}
{\cal E}(\mathbf n) = \lim_{r \to \infty} r^{d-3}\! \int_{-\infty}^\infty\! dt\; n^i\, T^0_{~i}(t, r\, \mathbf n) ~,
\label{etheta}
\end{equation}
the unit vector $\mathbf n$ pointing towards the actual direction of measure. The expectation value of the energy on a state created by a given local gauge invariant operator $\mathcal{O}$ is given by
\begin{equation}
\langle {\cal E}(\mathbf n) \rangle_\mathcal{O} = \frac{\langle 0| \mathcal{O}^\dagger {\cal E}(\mathbf n) \mathcal{O} |0 \rangle}{\langle 0| \mathcal{O}^\dagger \mathcal{O} |0 \rangle} ~.
\label{vevenergy}
\end{equation}
Thus, if $\mathcal{O} = \epsilon_{ij}\, T_{ij}$, $\langle {\cal E}(\mathbf n) \rangle_\mathcal{O}$ will be given in terms of 2- and 3-point correlators of $T_{\mu\nu}$, and rotational symmetry constraints it to be of the form
\begin{eqnarray}
\langle \mathcal E(\mathbf n) \rangle_\mathcal{O} &=& \frac{E}{\Omega_{d-3}}\left[1 + t_2 \left(\frac{n_i n_j \epsilon^*_{ik}\epsilon_{jk}}{\epsilon^*_{ik}\epsilon_{ik}}-\frac 1{d-2}\right) \qquad \right. \nonumber\\ [0.5em]
&& \left. \qquad\qquad\qquad + \ t_4 \left(\frac{|n_i n_j\epsilon_{ij}|^2}{\epsilon^*_{ik}\epsilon_{ik}}-\frac 2{d(d-2)}\right)\right] ~.
\label{flux}
\end{eqnarray}
$E$ is the total energy of the insertion and $\Omega_{d-3}$ the volume of a unit $(d-3)$-sphere. We have used the fact that numerator and denominator are quadratic in the polarization tensor, fixing the numerical coefficients inside the brackets so that the integration of $\langle \mathcal E(\mathbf n) \rangle_\mathcal{O}$ over the $(d-3)$-sphere gives the total energy. The energy flux is almost completely fixed by symmetry up to the coefficients $t_2$ and $t_4$, and since it is a quotient of 2- and 3-point functions of stress-energy tensor components, these coefficients should be writable in terms of the three parameters $\mathcal{A}$, $\mathcal{B}$ and $\mathcal{C}$. This was done in \cite{Buchel2010a} yielding
\begin{eqnarray}
t_2&=&\frac{2d}{d-1}\frac{(d-3)(d+1)d\mathcal{A}+3(d-1)^2\mathcal{B}-4(d-1)(2d-1)\mathcal{C}}{(d-2)(d+1)\mathcal{A}-2\mathcal{B}-4d\mathcal{C}} ~, \nonumber \\ [0.5em]
t_4&=&-\frac{d}{d-1}\bigg(\frac{(d+1)(2(d-1)^2-3(d-1)-3)\mathcal{A}}{(d-2)(d+1)\mathcal{A}-2\mathcal{B}-4d\mathcal{C}} ~\nonumber \\
& &\qquad \qquad \qquad +\frac{2(d-1)^2(d+1)\mathcal{B}-4(d-1)d(d+1)\mathcal{C}}{(d-2)(d+1)\mathcal{A}-2\mathcal{B}-4d\mathcal{C}} \bigg) ~.
\label{t2t4ABC}
\end{eqnarray}
Notice that \reef{flux} contains minus signs and, furthermore, the coefficients $t_2$ and $t_4$ are not necessarily positive. Thus, the above formulas seem compatible with an energy flux that is not manifestly positive definite.

\subsection{Constraints from positivity of the energy}

\begin{figure}[h]
\centering
\includegraphics[width=0.85\textwidth]{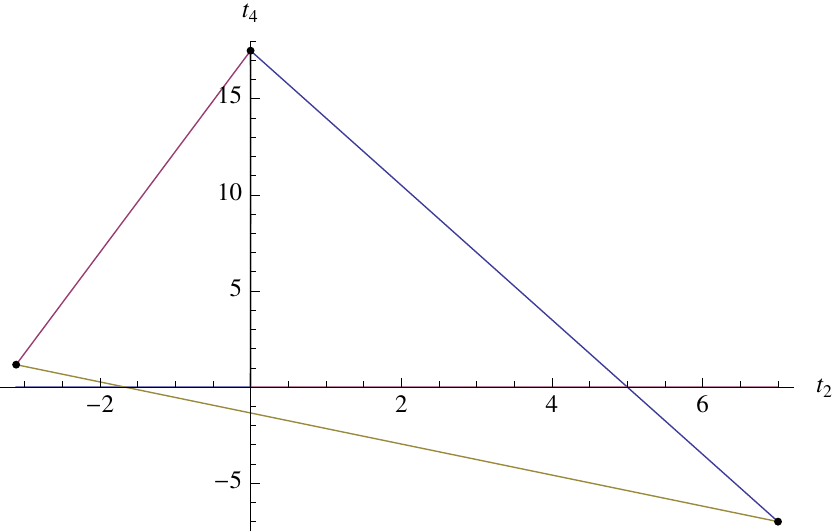}
\caption{Constraints (\ref{tensorboundd})--(\ref{scalarboundd}) restrict the values of $t_2$ and $t_4$ to the interior of a triangle with vertices in $(-\frac{2 (d-3) d}{d^2-5 d+4},\frac{d}{d-1})$, $(0,\frac{d(d-2)}{2})$ and $(d,-d)$; it is depicted, for definiteness, in the $d=7$ case.}
\label{triangle}
\end{figure}

The positivity of the energy flux for any direction $\mathbf n$ and polarization $\epsilon_{ij}$ seems to be a physically reasonable constraint on a well-defined CFT. Indeed, it holds in all known examples, and an almost complete proof of positivity can be attained \cite{Hofman2008,Hofman2009} (see also \cite{Zhiboedov} for a more recent discussion). It leads to constraints on the parameters $t_2$ and $t_4$, depending upon the splitting of $\epsilon_{ij}$ into tensor, vector and scalar components with respect to rotations in the plane perpendicular to $\mathbf n$,
\begin{eqnarray}
{\rm tensor:}\qquad & & 1 - \frac{1}{d-2}\, t_2 - \frac{2}{d(d-2)}\, t_4 \geq 0 ~,\label{tensorboundd} \\ [0.5em]
{\rm vector:}\qquad & & \left( 1 - \frac{1}{d-2}\, t_2 - \frac{2}{d(d-2)}\, t_4 \right) + \frac{1}{2}\, t_2 \geq 0 ~, \label{vectorboundd} \\ [0.5em]
{\rm scalar:}\qquad & & \left( 1 - \frac{1}{d-2}\, t_2 - \frac{2}{d(d-2)}\, t_4 \right) + \frac{d-3}{d-2} \left( t_2 + t_4 \right) \geq 0 ~. \label{scalarboundd}
\end{eqnarray}
These constraints restrict the possible values of $t_2$ and $t_4$ for any CFT, in arbitrary dimensions, to lie inside a triangle whose sides are given by (\ref{tensorboundd})--(\ref{scalarboundd}); see Figure \ref{triangle}.
Notice that this severe restriction does not require any {\it a priori} knowledge about the CFT, such as if its a Lagrangian theory, what are the relevant degrees of freedom, etc. Nevertheless, each of the constraints is saturated in a free theory with, respectively, no antisymmetric tensor fields, no fermions or no scalars \cite{Hofman2008}. Looking at the triangle, it is straightforward to see that the helicity one contribution is not restrictive for $t_4 < \frac{d}{d-1}$. In particular, this is the case for $t_4 = 0$.

The scalar, vector and tensor constraints coming from the positivity of energy can also be written, using (\ref{t2t4ABC}), in terms of the parameters $\mathcal{A}, \mathcal{B}$ and $\mathcal{C}$, leading to the following expressions:
\begin{eqnarray}
\frac{(d-3)(d+1)\mathcal{A}+2(d-1)\mathcal{B}-4(d-1)\mathcal{C}}{(d-2)(d+1)\mathcal{A}-2\mathcal{B}-4d\mathcal{C}} &\leq& 0 ~,\nonumber\\[0.5em]
\frac{(d-3)(d+1)\mathcal{A}+(3d-5)\mathcal{B}-8(d-1)\mathcal{C}}{(d-2)(d+1)\mathcal{A}-2\mathcal{B}-4d\mathcal{C}} &\geq& 0 ~,\\[0.5em]
\frac{\mathcal{B}-2\mathcal{C}}{(d-2)(d+1)\mathcal{A}-2\mathcal{B}-4d\mathcal{C}} &\leq& 0 ~.\nonumber
\end{eqnarray} 
If the CFT$_{d-1}$ is supersymmetric, $t_4$ vanishes. Evidence for this claim has been given in CFT$_4$ \cite{Hofman2008} and CFT$_6$ \cite{KulaxiziP1}. On the other hand, even though there is no proof in the literature showing that Lovelock theories admit a supersymmetric extension, it turns out that the holographic computation suggests that a CFT$_{d-1}$ with a weakly curved gravitational dual whose dynamics is governed by Lovelock theory will have a null value of $t_4$ \cite{Boer2010,Camanho2010a}. To the best of our knowledge, a supersymmetric extension of Lovelock theory has only been accomplished in the case of LGB theory in five dimensions \cite{Ozkan}. Inserting $t_4 = 0$ in \reef{flux}, we get
\begin{equation}
\langle \mathcal E(\mathbf n)\rangle = \frac{E}{\Omega_{d-3}} \left[ 1 + t_2 \left(\frac{n_i n_j \epsilon^*_{ik}\epsilon_{jk}}{\epsilon^*_{ik}\epsilon_{ik}}-\frac 1{d-2}\right) \right] ~.
\label{vevEfinal}
\end{equation}
Demanding positivity on the energy flux for any direction $\mathbf n$ and polarization $\epsilon_{ij}$, leads to a series of constraints on $t_2$. For the tensor, vector and scalar channels, we obtain, respectively,
\begin{equation}
t_2 \leq d-2 ~, \qquad t_2 \geq - \frac{2 (d-2)}{d-4} ~, \qquad t_2 \geq - \frac{d-2}{d-4} ~. 
\label{t2channels}
\end{equation}
As argued above, the vector channel constraint becomes irrelevant. In any Lovelock gravity dual to a CFT$_{d-1}$, therefore, $t_2$ has to take values within the window 
\begin{equation}
- \frac{d-2}{d-4} \leq t_2 \leq d-2 ~, 
\label{t2window}
\end{equation}
For instance, any $\mathcal{N}=1$ supersymmetric CFT$_4$ has $|t_2| \leq 3$, with
\begin{equation}
t_2 = 6\,\frac{c-a}{c} \qquad \Rightarrow \qquad \frac{1}{2} \leq \frac{a}{c} \leq \frac{3}{2} ~.
\label{t2ac}
\end{equation}
where $a$ and $c$ are the parameters entering the trace anomaly formula, the bound being saturated for free theories \cite{Hofman2008}.

\subsection{Holographic calculation of $t_2$ and $t_4$}

For conformal field theories with a weakly curved gravitational dual, it is possible to compute $t_2$ and $t_4$ holographically \cite{Camanho2010d}. The calculation proceeds by considering the vacuum AdS solution perturbed by a shock wave, which corresponds holographically to a $T_{--}$ insertion. By adding a transverse metric fluctuation, one reads off the interaction vertex from the action, and from that one obtains $t_2$ and $t_4$. Shock wave backgrounds in Lovelock theory were considered in \cite{Hofman2009,Camanho2010}, where it was found that in the presence of the shock wave there is room for causality violation in the dual field theory. The proviso that causality must hold in physically sensible quantum field theories places bounds on the Lovelock gravitational couplings which precisely match those portrayed in \reef{tensorboundd}--\reef{scalarboundd}.

Let us consider, along the lines of \cite{Camanho2010}, a helicity two perturbation $\phi(u,v,z)$ in the shock wave background (\ref{sw}),
\begin{equation}
d\tilde s^2_{{\rm AdS},sw} = ds^2_{{\rm AdS},sw} - \frac{2 \epsilon}{\Lambda_\star z^2}\, \phi(u,v,z)\,dx^2\,dx^3 ~.
\label{dsSW}
\end{equation}
This amounts to choosing just one non-vanishing component of the polarization tensor, $\epsilon_{23} \neq 0$. Leading contributions to the equations of motion, in the high momentum limit, come from the exterior derivative of the perturbation in the spin connection. The relevant equation of motion is $\delta \mathcal{E}_3\wedge e^3 = 0$, which, after some lengthy algebra, can be written as \cite{Camanho2010a}
\begin{equation}
\partial_u \partial_v \phi - \Lambda_\star z^2\, f(u)\,\varpi\, \left (1 - \frac{\Lambda_\star\,\Upsilon''[\Lambda_\star]}{\Upsilon'[\Lambda_\star]}\frac{T_2}{(d-3)(d-4)} \right) \partial_v^2 \phi = 0 ~,
\label{eq3}
\end{equation}
where 
\begin{equation}
T_2 = \frac{z^2(\partial_2^2 \varpi+\partial_3^2 \varpi)-2 z \partial_z \varpi-4 \varpi}{\varpi} ~.
\end{equation}
This is nothing but the same $T_2$ appearing in \cite{Buchel2010a}. There is an overall factor $\Upsilon'[\Lambda_\star]$ multiplying \reef{eq3}, as the reader may have expected from our earlier discussion regarding the unitarity properties of gravitational perturbations. For the shock wave profile, we shall consider a solution of the form (\ref{wsol3}). Such a profile has been argued in \cite{Hofman2008} to be the dual field configuration to $\mathcal{E}(\mathbf n)$ provided
\begin{equation*}
x_0^i=\frac{n^i}{1+n^{d-2}} ~, \qquad {\rm and} \qquad f(u)=\delta(u) ~.
\end{equation*}
We shall only focus on those terms proportional to $\partial_{v}^2\phi$ in \reef{eq3}. The $3$-point function follows from evaluating the effective action for the field $\phi$ on-shell, on a particular solution which depends on all coordinates, including $x^2$ and $x^3$ \cite{Buchel2010a}. The cubic interaction vertex of $\phi$ with the shock wave appearing in the action will be essentially the one in the equation of motion determined above. Up to an overall factor, the cubic vertex is then
\begin{equation}
\mathcal{I}^{(3)}\sim C_T \int dz\,d\mathbf x\, \sqrt{-g} \,\phi\, \partial^2_v \phi\, \varpi\,\left (1 - \frac{\Lambda_\star\,\Upsilon''[\Lambda_\star]}{\Upsilon'[\Lambda_\star]}\frac{T_2}{(d-3)(d-4)} \right) ~.
\end{equation}
Indeed, following \cite{Buchel2010a}, the relevant graviton profile is
\begin{equation}
\phi(z,u=0,v,\mathtt x) \sim e^{-iEv}\,\delta^{d-3}(\mathtt x)\,\delta(z-1) ~,
\end{equation}
so that we need to impose $\mathtt x = 0$ and $z=1$ yielding
\begin{equation}
T_2 = 2(d-1)(d-2)\left(\frac{n_2^2+n_3^2}{2}-\frac1{d-2}\right)~,
\end{equation}
and we therefore read off
\begin{equation}
t_2 = -\frac{2(d-1)(d-2)}{(d-3)(d-4)} \frac{\Lambda_\star \Upsilon''[\Lambda_\star]}{\Upsilon'[\Lambda_\star]} ~, \quad\qquad t_4 = 0 ~. 
\label{t2t4}
\end{equation}
As announced above, the holographic value of $t_4$ vanishes in Lovelock theory. Combining (\ref{t2window}) and (\ref{t2t4}), in turn, we obtain \cite{Boer2010,Camanho2010a}
\begin{equation}
- \frac{d-2}{d-4} \leq -\frac{2(d-1)(d-2)}{(d-3)(d-4)} \frac{{\Lambda_\star}\,\Upsilon''[\Lambda_\star]}{\Upsilon'[\Lambda_\star]} \leq d-2 ~.
\label{LLwindow}
\end{equation}
If specialized to the case of LGB gravity, equation (\ref{t2t4}) reproduces the results obtained in \cite{BuchelM,Boer2009,Camanho2010,Buchel2010a}. For the general case, it is exactly the same as conjectured in \cite{Boer2010,Camanho2010a}. Using these results altogether, including the holographic expression for $C_T$, we find formulas for the usual $3$-point function parameters $\mathcal A,\mathcal B, \mathcal C$, in terms of the Lovelock couplings:
\begin{eqnarray}
\mathcal{A} &=& - \frac{(d-1)^3}{(d-2)^3}\, \frac{\Gamma[d]}{\pi^{d-1}}\, \frac{1}{(-\Lambda_\star)^{d/2}} \left(\frac{2  d \Lambda_\star \Upsilon''[\Lambda_\star]}{(d-4)^2}+ \Upsilon'[\Lambda_\star]\right) ~, \nonumber \\[0.5em]
\mathcal{B} &=& -\frac{  (d-1) }{(d-2)^3 }\frac{\Gamma[d]}{\pi^{d-1} } \frac{1}{(-\Lambda_\star)^{d/2}}
\bigg(\frac{ (d-1) d \left(d^2-4 d + 6\right) \Lambda_\star \Upsilon''[\Lambda_\star]}{(d-4)^2} \nonumber \\
&& \qquad + \left(d^3-4d^2+5 d-1\right) \Upsilon'[\Lambda_\star]\bigg) ~, \nonumber\\ [0.5em]
\mathcal{C} &=& -\frac{(d-1)^2 }{2(d-2)^3}\frac{\Gamma[d]}{\pi^{d-1}} \frac{1}{(-\Lambda_\star)^{d/2}}
\bigg( \frac{ \left(d^3-3 d^2+3 d-4\right) \Lambda_\star \Upsilon''[\Lambda_\star]}{(d-4)^2} \nonumber \\
&& \qquad +\frac{1}{2} (2 (d-3) d+3)\, \Upsilon'[\Lambda_\star]\bigg) ~. \nonumber
\end{eqnarray}
In the coming section, we will show that the region of Lovelock parameters given by \reef{LLwindow} exactly matches the conditions we must impose to avoid causality violation.

\section{Causality violation in Lovelock theory}
\label{causality}

\subsection{Black hole perturbations}

Constraints coming from positivity of the energy will be shown to agree with those coming from imposing causality at the boundary theory due to the bulk gravity background. One way of checking this is by looking at perturbations of Lovelock black holes dual to thermal states of finite temperature CFTs. A detailed study of maximally symmetric black holes in Lovelock theory\footnote{Many important features of the static spherically symmetric solutions of Lovelock gravities were already understood in the late eighties \cite{Wheeler1986,Wheeler1986a,Myers1988,Wiltshire,Whitt}, greatly contributing to the acceptance of these theories as physically relevant. Subsequent work exploring in detail the case of degenerate Lovelock theory, {\it i.e.}, when the gravitational couplings are such that there is a unique (A)dS vacuum, have been pursued in \cite{BTZ1994,Crisostomo2000}. See also \cite{Charmousis,GG} for a nice recent report on the subject.} has been carried out in \cite{CamanhoEdelstein3}. The interested reader shall find all relevant formulas in that reference.

Having the holographic picture in mind, it is interesting to scrutinize the possibility, for these backgrounds, of having trajectories that start from the boundary of AdS and come back to it. These can be interpreted as bulk disturbances created by local operators in the boundary CFT, and we expect micro-causality violation in this theory if there exists a bouncing graviton traveling faster than light from the point of view of the boundary theory. This phenomenon may happen due to the fact that, in higher curvature gravity, gravitons do not propagate according to their background metric but, instead, feel an {\it effective metric} related to their equations of motion \cite{Brigante2008,Brigante2008a,Camanho2010}. In the large momentum limit, localized wave packets moving along null geodesics of this effective geometry satisfy radial equations of the form ($r = L^2/z$)
\begin{equation}
\left(\frac{dr}{d\tilde{s}}\right)^2 = \alpha^2 - {\bf c}_h^2(r) ~, \qquad \alpha\equiv\frac{\omega}{q} ~,
\end{equation}
equivalent to those of a particle of energy $\alpha^2$ moving in a potential given by ${\bf c}_h^2(r)$, which corresponds to the velocity of high momentum gravitons of helicity $h=0,1,2$ in the different radial slices. These potentials always go to one at the boundary, and they approach zero at the black hole horizon. In most cases, the potential is monotonic and, thereby, the graviton inevitably falls into the black hole. Whenever there is a maximum in ${\bf c}_h^2(r)$, in turn, geodesics starting at the boundary can be seen to find their way back to it, with turning point $\alpha^2 = {\bf c}_h^2(r_{\rm turn})$. For a null bouncing geodesic starting and ending at the boundary, as the energy $\alpha$ approaches the value of the speed at the maximum, $\alpha\rightarrow c_{2,{\rm max}}$ ({\it i.e.}, $r_{\rm turn} \rightarrow r_{\rm max}$), we have
\begin{equation}
\frac{\Delta x^{d-1}}{\Delta t} \rightarrow {\bf c}_{h,{\rm max}} > 1 ~.
\label{superlum}
\end{equation}
These geodesics spend an arbitrarily long time near the maximum, traveling with an average speed which is bigger than one. Interpreting this as originating from local operators in the boundary CFT, the hypothetical dual field theory will not be causal if there exists a bouncing geodesic obeying \reef{superlum}. In order to avert causality violation, we must demand these effective potentials to be always smaller than one \cite{Brigante2008,Brigante2008a}. Given that, in particular, at the boundary we have ${\bf c}_h^2=1$, we must demand $\partial_r {\bf c}_h^2 \geq 0$, as $r \to \infty$. This leads to the following constraints:
\begin{eqnarray}
{\rm Tensor:}\qquad & & \Upsilon'[\Lambda_\star]+\frac{2 (d-1)}{(d-3)(d-4)}\,\Lambda_\star \Upsilon''[\Lambda_\star] \geq 0 ~, \nonumber\\[0.5em]
{\rm Vector:}\qquad & & \Upsilon'[\Lambda_\star]-\frac{ (d-1)}{(d-3)}\,\Lambda_\star \Upsilon''[\Lambda_\star] \geq 0 ~, \label{genconstraints} \\[0.5em]
{\rm Scalar:}\qquad & & \Upsilon'[\Lambda_\star]-\frac{2 (d-1)}{(d-3)}\,\Lambda_\star \Upsilon''[\Lambda_\star] \geq 0 ~. \nonumber
\end{eqnarray}
for the three channels discussed above. These can be rewritten in terms of the dual CFT parameters, using the expressions for $t_2, t_4$ in \reef{t2t4}. Strikingly enough, the result is exactly \reef{tensorboundd}--\reef{scalarboundd} with $t_4=0$; that is, \reef{t2channels}.
\begin{figure}[h]
\centering
\includegraphics[width=0.94\textwidth]{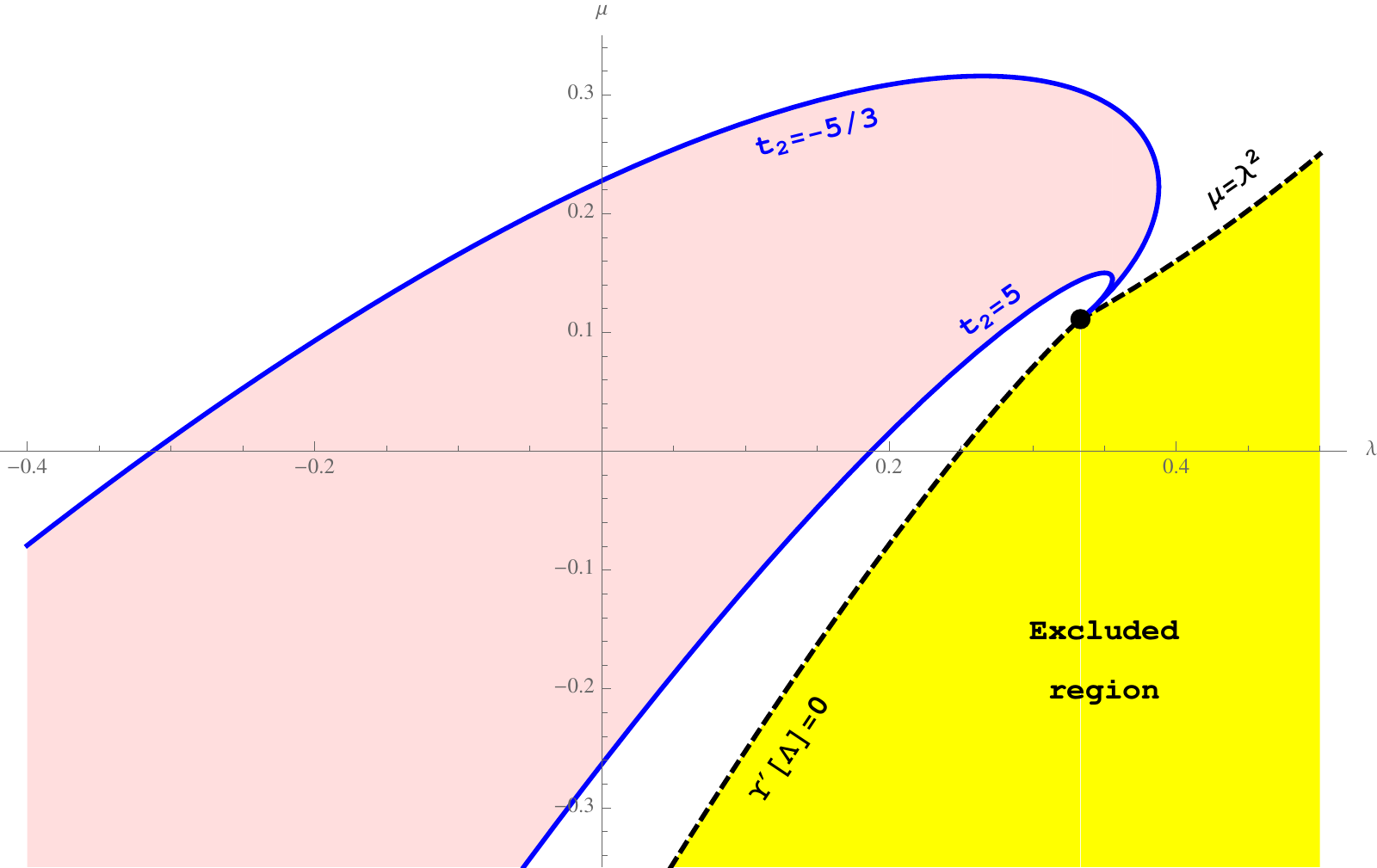}
\caption{The allowed window \reef{t2window} in seven dimensions becomes, after plugging in the holographic expressions for $t_2$ and $t_4$, the shadowed (pink) region with the shape of an eagle head. The gravitational couplings of Lovelock theory are constrained to belong to that region, in the AdS/CFT framework.}
\label{causal}
\end{figure}
In the seven dimensional case, the allowed window \reef{t2window} reads $-5/3 \leq t_2 \leq 5$. By means of \reef{t2t4} this can be translated into two curves that delimit the region of the space of Lovelock couplings compatible with causality; see Figure \ref{causal}. These are, of course, the tensor and scalar channel inequalities displayed in \reef{genconstraints}. In this way, the constraints posed by causality fully match those arising from the requirement of positivity of the energy in the dual conformal field theory.

\subsection{Scattering of gravitons and shock waves in AdS}

The previous computations are carried on a black hole background. As such, they are adequate in the context of thermal CFTs. As pointed out in \cite{Hofman2009}, causality violation should not be associated to a thermal feature, thereby one would expect to be able to perform a similar computation in a zero temperature background. An adequate background to perform a computation that is independent of the temperature is given by a pp-wave. In particular, it is easier to consider the simplest case, provided by shock waves \cite{Hofman2009}, since they are not subjected to higher derivative corrections \cite{Horowitz1999a}. As such, AdS shock waves are exact solutions in Lovelock theory (in string theory as well).

We will study the scattering of a graviton with an AdS shock wave in Lovelock theory. This computation, originally carried out by Hofman in the case of LGB gravity in 5d \cite{Hofman2009} (see also \cite{Lang}), and later extended to arbitrary higher dimensional spacetime \cite{Camanho2010}, can also be generalized to the case of any Lovelock theory \cite{Camanho2010a}. This process is, in a sense, the gravity dual of the energy 1-point function in the CFT \cite{Hofman2008}. We will see, once again, that causality violation poses a constraint on the allowed values of $t_2$. For forbidden values of this parameter, a graviton that is emitted from the boundary would come back and {\it land} outside its own light cone. The splitting of the graviton into different helicities will fully agree with the various polarization of the operator $\mathcal{O} = \epsilon_{ij}\, T_{ij}$ in \reef{flux}, for reasons that should be clear at this point of our discussion.

For definiteness, we present the computation in the helicity two channel. This amounts to the line element already given in \reef{dsSW}. The solution we are going to consider for the shock wave propagating on AdS is of the type displayed in the left expression in (\ref{wsol2}), which, as discussed in \cite{Hofman2009}, can be obtained from the black hole background by boosting the solution while keeping its energy constant. The normalization constant $\varpi_0$ is proportional to the energy density and, as such, must be positive if the original black hole solution had a positive mass.

We will compute the time delay, $\Delta v$, due to the collision of our perturbation with the shock wave, in order to analyze the occurrence (or not) of causality violation from the boundary point of view. For that it will be important  to make certain that the delay due to free propagation in AdS is negligible compared to the shock wave contribution. We will then need to consider the large momentum regime for our perturbation, in accordance with the analogous computation performed in the Lovelock black hole background. In this limit, the free propagation of a localized wave packet can be well approximated by geodesic motion in AdS that then yields
\begin{equation}
\Delta v^{\rm free} = 2\sqrt{\frac{P_u}{P_v}}\;z_\star ~,
\end{equation}
where $z_\star$ is the radial position of the collision point. We neglected the graviton motion in the transverse directions, thereby we see that we need $P_v \gg P_u$ (that also implies $P_v \gg P_z$). We only keep contributions of the sort $\partial^2_v\phi$ and $\partial_u\partial_v\phi$ in the equations of motion, as shown earlier in \reef{eq3}. The latter, even if subdominant, has to be kept to provide the dynamics of the graviton outside the locus of the shock wave.

Inserting $\varpi(\mathtt{x},z) = \varpi_0\, {z^{d-3}}$ in \reef{eq3}, we get
\begin{equation}
\partial_u\partial_v\phi - \varpi_0\,\Lambda_\star f(u) \ z^{d-1}\,\mathcal{N}_2\; \partial_v^2\phi = 0 ~,
\end{equation}
where $\mathcal{N}_2$ can be written in terms of $t_2$ defined in \reef{t2t4},
\begin{equation}
\mathcal{N}_2 = 1 - \frac{1}{d-2}\,t_2 ~.
\end{equation}
The computation for the other two helicities is harder, but the result is alike, with $\mathcal{N}_2$ replaced by $\mathcal{N}_h$, and
\begin{equation}
\mathcal{N}_1 = 1 + \frac{d-4}{2(d-2)}\,t_2 ~, \qquad\quad \mathcal{N}_0 = 1 + \frac{d-4}{d-2}\,t_2 ~.
\end{equation}
Taking the shock wave profile to be a delta function, $f(u) = \delta(u)$, the equation of motion reduces to the usual wave equation $\partial_u \partial_v \phi = 0$ outside the locus $u = 0$. Then, we can consider a wave packet moving with definite momentum on both sides of the shock wave. We can find a matching condition just by integrating the corresponding equation of motion along the discontinuity,
\begin{equation}
\phi_> = \phi_<\, e^{i P_v\, \varpi_0\,\Lambda_\star z^{d-1}\,\mathcal{N}_h} ~,
\label{matching}
\end{equation}
where $\phi_>$ and $\phi_<$ are the values of the perturbation at both sides of the discontinuity, and we used $P_v = - i \partial_v$. We can find the shift in the momentum in the $z$-direction acting with $P_z = - i \partial_z$,
\begin{equation}
P_z^> = P_z^< + (d-1) P_v\, \varpi_0\,\Lambda_\star z^{d-2}\,\mathcal{N}_h ~.
\end{equation}
If we consider a particle going inside AdS, $P_z^< > 0$. The momentum in the radial direction will change sign --the perturbation coming back to the boundary after the collision--, provided
\begin{equation}
P_v\, \varpi_0\,\Lambda_\star z^{d-2}\,\mathcal{N}_h < 0 ~,
\end{equation}
for sufficiently large $\varpi_0 > 0$ (since the black hole originating the shock wave had positive mass).

Now, both $P_v$ and $\Lambda_\star$ are negative, the former simply due to the fact that $P_v = - \frac12 P^u$ (and $P^u = P^0 + P^{d-1}$ must be positive for the energy to be so). Therefore, when
\begin{equation}
\mathcal{N}_h< 0 ~,
\end{equation}
for all three helicities, the graviton will make its way back to the boundary and, as we can thoroughly read from (\ref{matching}),
\begin{figure}[h]
\centering
\includegraphics[width=0.43\textwidth]{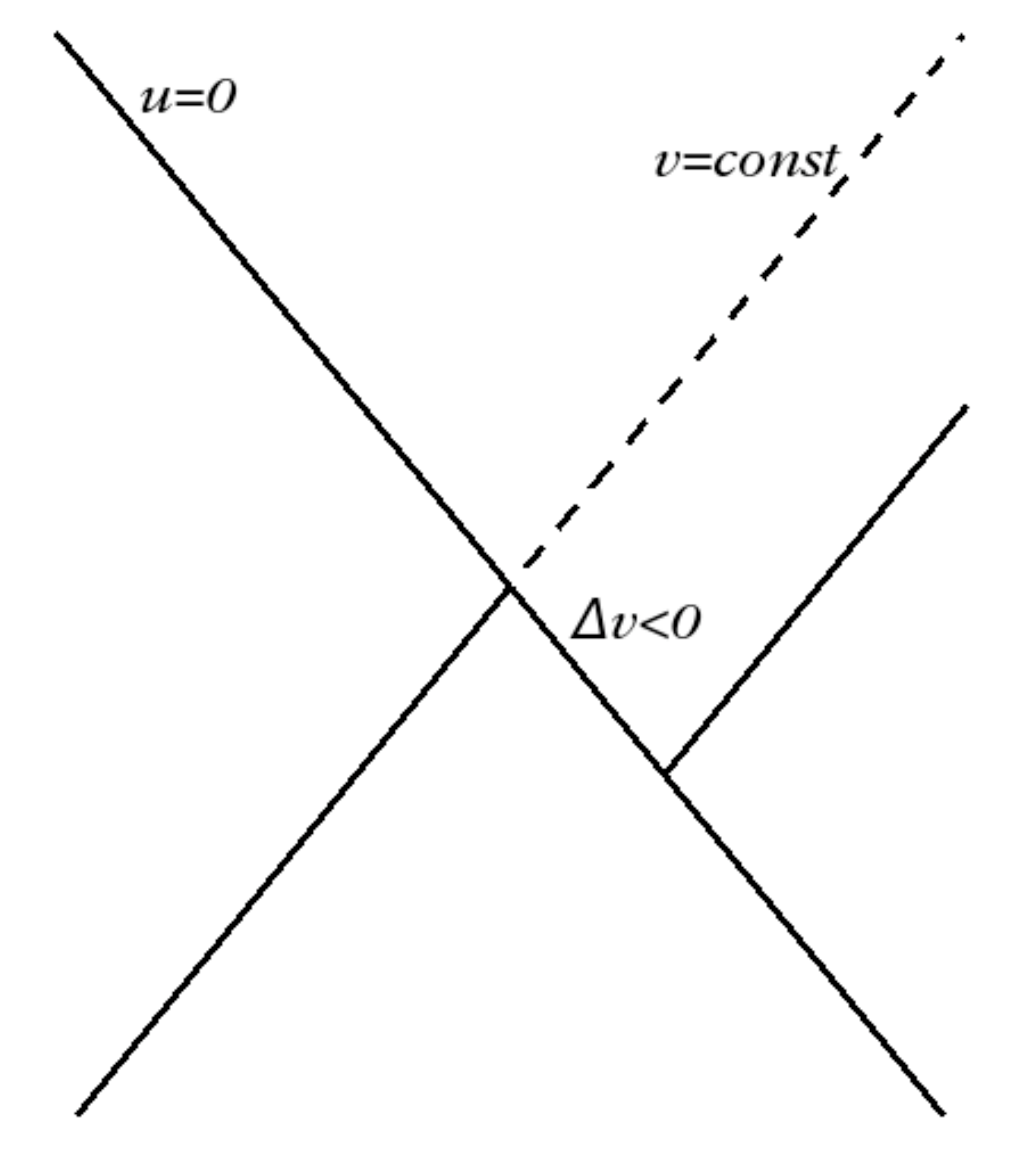}\caption{The line $u=0$ corresponds to the shock wave while the line $v=const.$ corresponds to the graviton. After the collision, if $\Delta v<0$, the particle lands outside its light-cone.}\label{shockwave}
\end{figure}
it comes back shifted in the $v$-direction a negative amount (see figure \ref{shockwave})
\begin{equation}
\Delta v = - \varpi_0\,\Lambda_\star z^{d-1}\, \mathcal{N}_h ~.
\end{equation}
The graviton lands, at the boundary, outside its own light-cone. This is an explicit break up of causality. We conclude that the theory violates causality unless $\mathcal{N}_h \geq 0$ for the three helicity channels, which amounts exactly to the same constraints found in the black hole case and, as well, to those arising from positivity of the energy in the dual CFT, displayed in \reef{t2channels}.

\section{Final comments and conclusions}
\label{final}

As we have seen throughout these pages, Lovelock theories of gravity appear as a very useful playground in order to analyze several aspects of the gauge/gravity duality. The precise status of these theories is at present unclear, as they do not generally appear in the low energy action of string theory, at least not with finite coefficients, and their ultraviolet completion is unknown. Although this is true, Lovelock gravities are two derivative theories, the most natural extension of General Relativity in higher dimensions, and there is no a priori reason why these cannot appear as classical limits of a quantum theory of gravity, thus subject to the holographic principle.

From the CFT point of view, moreover, the inclusion of these higher curvature terms allows for the description of more general field theories, a notable example being strongly coupled CFTs with different central charges, $a \neq c$, in four dimensions \cite{Nojiri} at leading order in the large N limit. Besides, the analysis of the would be CFT duals of Lovelock gravities has lead to some very interesting and unsuspected insights. One that has been extensively reviewed in this article is the relation between causality and positivity of the energy. Our analysis indicates that the violation of the positivity constraints leads to the appearance of superluminal modes propagating in the field theory. Using purely field theory techniques it has been shown that the same constraints are needed to avoid the presence of ghosts at finite temperature \cite{Kulaxizi2010}. 

Many efforts have been devoted in the last few years to the analysis of some of the preeminent features of Lovelock theory such as its black hole solutions, their thermodynamics and phase transitions, their would be instabilities and the generic existence of more than one maximally symmetric solution or {\it vacuum}. Even though it seems to us that the last word has not been said, the results obtained so far indicate that there is no obvious pathology that invalidates the consideration of these theories once and for all, at least for some (finite) region of the parameter space that includes the Einstein-Hilbert case. Some authors have pointed to some instabilities that quite generically appear in these theories as being a sign of their {\it sickness}. Much on the contrary, these instabilities have been found to play a central r\^ole in the dynamics of the theory. For instance, they generically prevent the formation of naked singularities thus being instrumental in the {\it cosmic censorship hypothesis} coming into being in this context\cite{Camanho2013b}.

In the present article we have focused our discussion in a few examples where Lovelock theory has been shown to yield interesting holographic connections between the gravitational and the field theory dynamics, connections that would otherwise be impossible to uncover in the simpler setup of General Relativity. Many more examples exist, though, of the convenience of this extended framework for the discussion of the most diverse issues. One such example concerns the existence of an analog of Zamolodchikov's $c$-theorem \cite{Zamolodchikov1986} in higher dimensions, namely for 4d CFTs. Even though some previous indications existed \cite{Jack1990}, the existence of a would be monotonic quantity and how this would be related to the central charges of the theory in four dimensions, $a$ and $c$, was not properly established. Under a holographic RG flow, the existence of a monotonic quantity related to the central charge $a$ has been pointed out in \cite{Myers2010}, much earlier than the actual field theoretic proof of the so-called $a$-theorem was found \cite{Komargodski2011}. No monotonicity property is known for the other central charge, $c$. The use of higher curvature theories is essential here as otherwise it would be impossible to identify which is the central charge with the monotonic property, given that $a$ and $c$ have the same value if gravity is governed by the Einstein-Hilbert action.

Another important issue that is receiving a lot of attention lately is that of entanglement entropy. The Ryu-Takayanagi proposal \cite{Ryu2006} for the holographic computation of this important quantity amounts to the determination of a minimal surface enclosing the region of interest at the boundary, the entanglement entropy being simply proportional to the area of such surface. This conjecture has recently been accounted for in \cite{Lewkowycz2013}. The na\"ive generalization of this prescription to higher curvature gravities would be that the entropy is given by the Wald formula, but this has been argued to be wrong \cite{Hung2011}. A more educated guess would be to consider the Wald-Iyer formula \cite{Iyer1994}, that reduces to the modified proposal in \cite{Hung2011} for Lovelock gravity \cite{deBoer2011}. The checks performed in order to find the discrepancies involve certain specific contributions to the entanglement entropy that are completely governed by the conformal anomaly. Depending on the shape of the region of interest, those contributions have to be proportional to a different combination of the central charges of the theory. The Wald proposal, instead, yields the same combination of central charges independently of the shape. In the case of Einstein-Hilbert gravity we have just one independent combination of the central charges, and this is the reason behind the fact that the simplified area prescription works.

Another obvious field where higher curvature gravity theories have shown their value and utility is that of the fluid/gravity correspondence (see, for instance, \cite{Hubeny2011} for a review of one possible approach). In this framework, similarly to the CFT case, Lovelock gravities allow for the description of relativistc fluid duals with more general transport coefficients than those resulting from the Einstein-Hilbert action. The most celebrated example is that of the shear viscosity to entropy density ratio, $\eta/s$ (see \cite{Cremonini2011} for a recent review, and references therein). In the context of higher curvature gravity theories, this specific transport coefficient has been considered in many papers (see \cite{Buchel2008b,Buchel2009,Buchel2009b,Buchel2010} for some relevant examples).\footnote{A different and interesting approach has been pursued in \cite{Hu2011}.}

Higher curvature theories, Lovelock in particular, have been used to disprove the longstanding KSS viscosity bound conjecture \cite{Kovtun2005}. The value of the shear viscosity to entropy density in general Lovelock theories has been established in \cite{Shu2009b}, where it was proven that the only Lovelock coefficient affecting the actual value of $\eta/s$ is the LGB one; any positive value of $\lambda$ translating into a violation of the viscosity bound. The causality/positivity constraints discussed in the core of this paper have been argued to impose a new bound for any Lovelock theory in any number of space-time dimensions, even though also stability constraints have to be considered to avoid negative values of $\eta$ in general \cite{Camanho2010d}. In spite of the fact that higher order Lovelock terms do not enter the holographic formula of the shear viscosity, they do change the actual value of the bound. A way of lifting the causality constraints on $\eta/s$ has been proposed in \cite{Buchel2010}, but stability still implies the existence of a finite bound. It is worth mentioning at this point that higher curvature corrections that are not of the Lovelock type are also relevant in this discussion as, for instance, in the case of quasi topological gravity where $t_4 \neq 0$ and the bound can be slightly lowered \cite{MyersPS}. In this respect, the Lovelock setup shall be viewed as an exploratory playground where computations are under better control due to the second order nature of its equations of motion.

Another possible connection between Lovelock theory and the fluid/gravity correspondence may have to do with the realization in \cite{Bhattacharya} that the existence of a positive divergence entropy current in an arbitrary
curved background and the Onsager principle put severe constraints in the terms appearing at first order in the hydrodynamics derivative expansion. It is natural to wonder whether similar constraints appear in gravity, restricting the space of allowed higher derivative corrections to Einstein's equations \cite{Minwalla}. This may be particularly relevant for those cases in which the entropy production vanishes in Einstein gravity, and one might have to go to the next correction to realize which is the sign of the divergence of the entropy current.

Surely many more applications of Lovelock theory in the framework of the AdS/CFT correspondence were left into the ink pot. It is a lively subject in which we expect to see further progress happening soon.

\begin{acknowledgements}
We wish to thank Alex Buchel, Gast\'on Giribet, Andy Gomberoff, Diego Hofman, Manuela Kulaxizi, Juan Maldacena, Rob Myers, Miguel Paulos and Sasha Zhiboedov for discussions on these subjects held over the last few years.
This work was supported in part by MICINN and FEDER (grant FPA2011-22594), by Xunta de Galicia (Conseller\'{\i}a de Educaci\'on and grant PGIDIT10PXIB206075PR), and by the Spanish Consolider-Ingenio 2010 Programme CPAN (CSD2007-00042).
X.O.C. is thankful to the Front of Galician-speaking Scientists for encouragement.
\end{acknowledgements}


\begin{thebibliography}{99}

\bibitem{Lovelock1971}
D.~Lovelock,
{\it The Einstein tensor and its generalizations}, 
{\em J.\ Math.\ Phys.}\ {\bf 12} (1971) 498.

\bibitem{Zwiebach}
B.~Zwiebach,
{\it Curvature squared terms and string theories},
{\em Phys.\ Lett.\ B} {\bf 156} (1985) 315.

\bibitem{Zumino1986}
B.~Zumino,
{\it Gravity theories in more than four dimensions}, 
{\em Phys.\ Rept.}\ {\bf 137} (1986) 109.

\bibitem{Maldacena}
J.~M.~Maldacena,
{\it The large N limit of superconformal field theories and supergravity},
{\em Adv.\ Theor.\ Math.\ Phys.}\ {\bf 2} (1998) 231.

\bibitem{Padmanabhan2013}
T.~Padmanabhan and D.~Kothawala,
{\it Lanczos-Lovelock models of gravity},
arXiv:1302.2151 [gr-qc].

\bibitem{Edelstein2013}
J.~D.~Edelstein,
{\it Lovelock theory, black holes and holography},
arXiv:1303.6213 [gr-qc].

\bibitem{Eguchi1980}
T.~Eguchi, P.~B.~Gilkey and A.~J.~Hanson, 
{\it Gravitation, gauge theories and differential geometry}, 
{\em Phys.\ Rept.}\ {\bf 66} (1980) 213.
  
\bibitem{Willison2004}
E.~Gravanis and S.~Willison,
{\it Intersecting membranes in AdS and Lovelock gravity},
{\em J.\ Math.\ Phys.}\ {\bf 47} (2006) 092503

\bibitem{Zanelli2005}
J.~Zanelli,
{\it Lecture notes on Chern-Simons (super-)gravities},
hep-th/0502193.

\bibitem{Regge}
T.~Regge,
{\it On broken symmetries and gravity},
{\em Phys.\ Rept.}\ {\bf 137} (1986) 31.

\bibitem{Lanczos}
C.~Lanczos,
{\it A Remarkable property of the Riemann-Christoffel tensor in four dimensions},
{\em Annals Math.}\ {\bf 39} (1938) 842.

\bibitem{Madore}
J.~Madore,
{\it Kaluza-Klein theory with the Lanczos lagrangian},
{\em Phys.\ Lett.\ A} {\bf 110} (1985) 289.

\bibitem{Gibbons1977}
G.~W.~Gibbons and S.~W.~Hawking, 
{\it Action integrals and partition functions in quantum gravity}, 
{\em Phys.\ Rev.\ D} {\bf 15} (1977) 2752.

\bibitem{Myers1987}
R.~Myers,
{\it Higher-derivative gravity, surface terms, and string theory},
{\em Phys.\ Rev.\ D} {\bf 36} (1987) 392.

\bibitem{Israel1967}
W.~Israel,
{\it Singular hypersurfaces and thin shells in general relativity},
{\em Nuovo Cim.\ B} {\bf 44} (1966) 1 [{\em Erratum-ibid.\ B} {\bf 48} (1967) 463].

\bibitem{Gravanis2003a}
E.~Gravanis and S.~Willison, 
{\it Israel conditions for the Gauss-Bonnet theory and the Friedmann equation on the brane universe},
{\em Phys.\ Lett.\ B} {\bf 562} (2003) 118.

\bibitem{CEGG1}
X.~O.~Camanho, J.~D.~Edelstein, G.~Giribet and A.~Gomberoff,
{\it New type of phase transition in gravitational theories},
{\em Phys.\ Rev.\ D} {\bf 86} (2012) 124048.

\bibitem{CEGG2}
X.~O.~Camanho, J.~D.~Edelstein, G.~Giribet and A.~Gomberoff,
{\it Generalized phase transition in Lovelock theory},
to appear, 2013.

\bibitem{Boulware1985a}
D.~G. Boulware and S.~Deser,
{\it String generated gravity models},
{\em Phys.\ Rev.\ Lett.}\ {\bf 55} (1985) 2656.

\bibitem{Horowitz1999}
G.~T.~Horowitz and N.~Itzhaki,
{\it Black holes, shock waves, and causality in the AdS/CFT correspondence},
{\em JHEP} {\bf 9902} (1999) 010.

\bibitem{Hofman2008}
D.~M.~Hofman and J.~Maldacena,
{\it Conformal collider physics: Energy and charge correlations},
{\em JHEP} {\bf 0805} (2008) 012.

\bibitem{Aharony1999}
O.~Aharony, S.~S.~Gubser, J.~M.~Maldacena, H.~Ooguri and Y.~Oz,
{\it Large N field theories, string theory and gravity},
{\em Phys.\ Rept.}\ {\bf 323} (2000) 183.

\bibitem{Edelstein2006}
J.~D.~Edelstein and R.~Portugues,
{\it Gauge/string duality in confining theories},
{\em Fortsch.\ Phys.}\ {\bf 54} (2006) 525.

\bibitem{Casalderrey2011}
J.~Casalderrey-Solana, H.~Liu, D.~Mateos, K.~Rajagopal and U.~A.~Wiedemann,
{\it Gauge/string duality, hot QCD and heavy ion collisions},
arXiv:1101.0618 [hep-th].

\bibitem{GKP}
S.~S.~Gubser, I.~R.~Klebanov and A.~M.~Polyakov, 
{\it Gauge theory correlators from non-critical string theory}, 
{\em Phys.\ Lett.\ B} {\bf 428} (1998) 105.

\bibitem{Witten98}
E.~Witten, 
{\it Anti-de Sitter space and holography}, 
{\em Adv.\ Theor.\ Math.\ Phys.}\ {\bf 2} (1998) 253.

\bibitem{OsbornPetkou} 
H.~Osborn and A.~C.~Petkou,
{\it Implications of conformal invariance in field theories for general dimensions},
{\em Annals Phys.}\ {\bf 231} (1994) 311.

\bibitem{Buchel2010a}
A.~Buchel, J.~Escobedo, R.~C. Myers, M.~F.~Paulos, A.~Sinha and M.~Smolkin,
{\it Holographic GB gravity in arbitrary dimensions}, 
{\em JHEP} {\bf 1003} (2010) 111.

\bibitem{Camanho2010d}
X.~O.~Camanho, J.~D.~Edelstein and M.~F.~Paulos, 
{\it Lovelock theories, holography and the fate of the viscosity bound}, 
{\em JHEP} {\bf 1105} (2011) 127.

\bibitem{Arutyunov1999}
G.~Arutyunov and S.~Frolov,
{\it Three point Green function of the stress energy tensor in the AdS/CFT correspondence},
{\em Phys.\ Rev.\ D} {\bf 60} (1999) 026004.

\bibitem{Liu1998}
H.~Liu and A.~A.~Tseytlin,
{\it D=4 super Yang-Mills, D=5 gauged supergravity, and D=4 conformal supergravity},
{\em Nucl.\ Phys.\ B} {\bf 533} (1998) 88.

\bibitem{Erdmenger1997}
J.~Erdmenger and H.~Osborn, 
{\it Conserved currents and the energy-momentum tensor in conformally invariant theories for general dimensions}, 
{\em Nucl.\ Phys.\ B} {\bf 483} (1997) 431.

\bibitem{Hofman2009}
D.~M.~Hofman, 
{\it Higher derivative gravity, causality and positivity of energy in a UV complete QFT}, 
{\em Nucl.\ Phys.\ B} {\bf 823} (2009) 174.

\bibitem{Zhiboedov}
A.~Zhiboedov,
{\it On conformal field theories with extremal a/c values},
arXiv:1304.6075 [hep-th].

\bibitem{KulaxiziP1} 
M.~Kulaxizi and A.~Parnachev,
{\it Supersymmetry constraints in holographic gravities},
{\em Phys.\ Rev.\ D} {\bf 82} (2010) 066001.

\bibitem{Boer2010}
J.~de~Boer, M.~Kulaxizi and A.~Parnachev, 
{\it Holographic Lovelock gravities and black holes}, 
{\em JHEP} {\bf 1006} (2010) 008.

\bibitem{Camanho2010a}
X.~O. Camanho and J.~D.~Edelstein, 
{\it Causality in AdS/CFT and Lovelock theory}, 
{\em JHEP} {\bf 1006} (2010) 099.

\bibitem{Ozkan}
 M.~Ozkan and Y.~Pang,
{\it Supersymmetric completion of Gauss-Bonnet combination in five dimensions}, 
{\em JHEP} {\bf 1303} (2013) 158.

\bibitem{Camanho2010}
X.~O. Camanho and J.~D. Edelstein, 
{\it Causality constraints in AdS/CFT from conformal collider physics and Gauss-Bonnet gravity}, 
{\em JHEP} {\bf 1004} (2010) 007.

\bibitem{BuchelM} 
A.~Buchel and R.~C.~Myers,
{\it Causality of holographic hydrodynamics},
{\em JHEP} {\bf 0908} (2009) 016.

\bibitem{Boer2009}
J.~de~Boer, M.~Kulaxizi and A.~Parnachev, 
{\it AdS$_7$/CFT$_6$, Gauss-Bonnet gravity, and viscosity bound}, 
{\em JHEP} {\bf 1003} (2010) 087.

\bibitem{Wheeler1986}
J.~T.~Wheeler,
{\it Symmetric solutions to the Gauss-Bonnet extended Einstein equations},
{\em Nucl.\ Phys.\ B} {\bf 268} (1986) 737.

\bibitem{Wheeler1986a}
J.~T.~Wheeler,
{\it Symmetric solutions to the maximally Gauss-Bonnet extended Einstein equations},
{\em Nucl.\ Phys.\ B} {\bf 273} (1986) 732.

\bibitem{Myers1988}
R.~C.~Myers and J.~Z.~Simon,
{\it Black Hole thermodynamics in Lovelock gravity},
{\em Phys.\ Rev.\ D} {\bf 38} (1988) 2434.

\bibitem{Wiltshire}
D.~L.~Wiltshire,
{\it Black holes in string generated gravity models},
{\em Phys.\ Rev.\ D} {\bf 38} (1988) 2445.

\bibitem{Whitt}
B.~Whitt,
{\it Spherically symmetric solutions of general second order gravity},
{\em Phys.\ Rev.\ D} {\bf 38} (1988) 3000.

\bibitem{BTZ1994}
M.~Ba\~nados, C.~Teitelboim and J.~Zanelli,
{\it Black hole entropy and the dimensional continuation of the Gauss-Bonnet theorem},
{\em Phys.\ Rev.\ Lett.}\ {\bf 72} (1994) 957.

\bibitem{Crisostomo2000}
J.~Crisostomo, R.~Troncoso and J.~Zanelli,
{\it Black hole scan},
{\em Phys.\ Rev.\ D} {\bf 62} (2000) 084013.

\bibitem{Charmousis}
C.~Charmousis,
{\it Higher order gravity theories and their black hole solutions},
{\em Lect.\ Notes Phys.}\ {\bf 769} (2009) 299.

\bibitem{GG}
C.~Garraffo and G.~Giribet,
{\it The Lovelock black holes},
{\em Mod.\ Phys.\ Lett.\ A} {\bf 23} (2008) 1801.

\bibitem{CamanhoEdelstein3}
X.~O.~Camanho and J.~D.~Edelstein,
{\it A Lovelock black hole bestiary},
{\em Class.\ Quant.\ Grav.}\ {\bf 30} (2013) 035009.

\bibitem{Brigante2008}
M.~Brigante, H.~Liu, R.~C.~Myers, S.~Shenker and S.~Yaida, 
{\it Viscosity bound violation in higher derivative gravity}, 
{\em Phys.\ Rev.\ D} {\bf 77} (2008) 126006.

\bibitem{Brigante2008a}
M.~Brigante, H.~Liu, R.~C.~Myers, S.~Shenker and S.~Yaida, 
{\it The viscosity bound and causality violation}, 
{\em Phys.\ Rev.\ Lett.}\ {\bf 100} (2008) 191601.

\bibitem{Horowitz1999a}
G.~T.~Horowitz and N.~Itzhaki, 
{Black holes, shock waves, and causality in the AdS/CFT correspondence},
{\em JHEP} {\bf 9902} (1999) 010.

\bibitem{Lang} 
R.~Lang,
{\it Propagation of gravitons in the shock wave geometry},
B.Sc. Thesis, MIT, 2009.
http://hdl.handle.net/1721.1/51580

\bibitem{Nojiri}
S.~'i.~Nojiri and S.~D.~Odintsov,
{\it On the conformal anomaly from higher derivative gravity in AdS/CFT correspondence},
{\em Int.\ J.\ Mod.\ Phys.\ A} {\bf 15} (2000) 413.

\bibitem{Kulaxizi2010}
M.~Kulaxizi and A.~Parnachev,
{\it Energy Flux Positivity and Unitarity in CFTs,}
{\em Phys.\ Rev.\ Lett.}\ {\bf 106} (2011) 011601.

\bibitem{Camanho2013b}
X.~O.~Camanho and J.~D.~Edelstein,
{\it Cosmic censorship in Lovelock theory},
arXiv:1308.0304 [hep-th].

\bibitem{Zamolodchikov1986}
A.~B.~Zamolodchikov,
{\it Irreversibility of the flux of the renormalization group in a 2D field theory},
{\em JETP Lett.}\ {\bf 43} (1986) 730 [{\em Pisma Zh.\ Eksp.\ Teor.\ Fiz.}\ {\bf 43} (1986) 565].

\bibitem{Jack1990}
I.~Jack and H.~Osborn,
{\it Analogs for the c-theorem for four-dimensional renormalizable field theories},
{\em Nucl.\ Phys.\ B} {\bf 343} (1990) 647.

\bibitem{Myers2010}
R.~C.~Myers and A.~Sinha,
{\it Holographic c-theorems in arbitrary dimensions},
{\em JHEP} {\bf 1101} (2011) 125.

\bibitem{Komargodski2011}
Z.~Komargodski and A.~Schwimmer,
{\it On renormalization group flows in four dimensions},
{\em JHEP} {\bf 1112} (2011) 099.

\bibitem{Ryu2006}
S.~Ryu and T.~Takayanagi,
{\it Holographic derivation of entanglement entropy from AdS/CFT},
{\em Phys.\ Rev.\ Lett.}\ {\bf 96} (2006) 181602.

\bibitem{Lewkowycz2013}
 A.~Lewkowycz and J.~Maldacena,
{\it Generalized gravitational entropy},
arXiv:1304.4926 [hep-th].

\bibitem{Hung2011}
L.~-Y.~Hung, R.~C.~Myers and M.~Smolkin,
{\it On holographic entanglement entropy and higher curvature gravity},
{\em JHEP} {\bf 1104} (2011) 025.

\bibitem{Iyer1994}
V.~Iyer and R.~M.~Wald,
{\it Some properties of Noether charge and a proposal for dynamical black hole entropy},
{\em Phys.\ Rev.\ D} {\bf 50} (1994) 846.

\bibitem{deBoer2011}
J.~de Boer, M.~Kulaxizi and A.~Parnachev,
{\it Holographic entanglement entropy in Lovelock gravities},
{\em JHEP} {\bf 1107} (2011) 109.

\bibitem{Hubeny2011}
V.~E.~Hubeny, S.~Minwalla and M.~Rangamani,
{\it The fluid/gravity correspondence},
arXiv:1107.5780 [hep-th].

\bibitem{Cremonini2011}
S.~Cremonini,
{\it The shear viscosity to entropy ratio: A status report},
{\em Mod.\ Phys.\ Lett.\ B} {\bf 25} (2011) 1867.

\bibitem{Buchel2008b}
A.~Buchel,
{\it Shear viscosity of CFT plasma at finite coupling},
{\em Phys.\ Lett.\ B} {\bf 665} (2008) 298.

\bibitem{Buchel2010}
A.~Buchel and S.~Cremonini,
{\it Viscosity bound and causality in superfluid plasma},
{\em JHEP} {\bf 1010} (2010) 026.

\bibitem{Buchel2009b}
A.~Buchel, M.~P.~Heller and R.~C.~Myers,
{\it sQGP as hCFT},
{\em Phys.\ Lett.\ B} {\bf 680} (2009) 521.

\bibitem{Buchel2009}
A.~Buchel, R.~C.~Myers and A.~Sinha,
{\it Beyond $\eta/s = 1/4\pi$},
{\em JHEP} {\bf 0903} (2009) 084.

\bibitem{Hu2011}
Y.~-P.~Hu, H.~-F.~Li and Z.~-Y.~Nie,
{\it The first order hydrodynamics via AdS/CFT correspondence in the Gauss-Bonnet gravity},
{\em JHEP} {\bf 1101} (2011) 123.

\bibitem{Kovtun2005}
P.~Kovtun, D.~T. Son, and A.~O.~Starinets,
{\it Viscosity in strongly interacting quantum field theories from black hole physics},
{\em Phys\. Rev.\ Lett.}\ {\bf 94} (2005) 111601.

\bibitem{Shu2009b}
F.-W.~Shu,
{\it The quantum viscosity bound in Lovelock gravity},
{\em Phys.\ Lett.\ B} {\bf 685} (2010) 325.

\bibitem{MyersPS}
R.~C.~Myers, M.~F.~Paulos and A.~Sinha,
{\it Holographic studies of quasi-topological gravity},
{\em JHEP} {\bf 1008} (2010) 035.

\bibitem{Bhattacharya}
J.~Bhattacharya, S.~Bhattacharyya, S.~Minwalla and A.~Yarom,
{\it A theory of first order dissipative superfluid dynamics},
arXiv:1105.3733 [hep-th].

\bibitem{Minwalla}
S.~Minwalla,
{\it The entropy current in hydrodynamics, superfluid hydrodynamics and gravity},
talk at Strings 2011.

\end{thebibliography}
\end{document}